\documentclass[aps,showpacs,preprintnumbers, amssymb,onecolumn,11pt,preprintnumbers,amsmath,amssymb,floatfix,prd,showpacs,superscriptaddress,nofootinbib,showkeys]{revtex4}

\usepackage{float}
\usepackage{graphics,epsfig}
\usepackage{graphicx}
\usepackage{dcolumn}
\usepackage{bm}
\begin{document}
\baselineskip=0.8 cm
\title{{\bf   Analytical    holographic superconductor with
backreaction using $AdS_3/CFT_2$}}
\author{Davood Momeni } \email{d.momeni@yahoo.com}\affiliation{Eurasian International Center for Theoretical Physics,
Eurasian National University, Astana 010008, Kazakhstan}
\author{ Muhammad Raza}\email{mreza06@gmail.com}\affiliation{Department of Computer Science, COMSATS Institute of
Information Technology, Sahiwal, Pakistan}

\author{Mohammad Reza Setare}
\affiliation{Department of Science, Campus of Bijar, University of Kurdistan Bijar, Iran}
\email{rezakord@ipm.ir}

\author{Ratbay Myrzakulov} \email{rmyrzakulov@gmail.com}  \affiliation{Eurasian International Center for Theoretical Physics,
Eurasian National University, Astana 010008, Kazakhstan}

\vspace*{0.2cm}
\begin{abstract}
\baselineskip=0.6 cm
\begin{center}
{\bf Abstract}
\end{center}

The holographic model for a two-dimensional superconductor has been
investigated by considering the three-dimensional gravity in the
bulk. To find the critical temperature, we used the Sturm-Liouville
variational method. Where as the same method is applied for
calculating the condensation of the dual operators on the boundary.
We included the back reactions on the metric by a combination of the
perturbation method of the fields with respect to the small
parameter and then applying the variational integrals on the
resulting equations of the motion. The critical temperature has been
successfully obtained on the backreaction effects, and we showed
that it dropped with a rise in the backreaction of the fields, and
it makes the condensation harder. We can use our analytical results
to support the numerical data which was reported previously.
\end{abstract}

\pacs{11.25.Tq, 04.70.Bw, 74.20.-z}
\keywords{Black Holes in String Theory, AdS/CFT Correspondence, Holography and
condensed matter physics (AdS/CMT)}
\maketitle
\newpage
\vspace*{0.2cm}

\section{Introduction}
Maldacenda discovered the relation between a $d+1$ dimensional
gravitational system as the weak limit of the string theory, and a
quantum theory on the $d$ dimensional boundary, which is illustrated
by the conformal field theory \cite{maldacena}. It is called anti-de
Sitter/conformal field theories (AdS/CFT) communication. In recent
years, its applications  has been frequently observed to explain the behaviors of 
the simple strongly correlated systems
\cite{HartnollRev,HerzogRev,HorowitzRev}in the theoretical condensed matter
physics \cite{HorowitzPRD78}-\cite{scripta} . A large number of the
holographic superconductors in a wide class of the gravitational
models,have been investigated from the non relativistic regime in
the form of Horava-Lifshitz to the Gauss-Bonnet and Weyl
corrections. If we ignore the field's backreaction effects on
gravity sector and fix the metric as static and with a definitive
symmetry and a well behavior horizon's temperature, this
approximation is called as the  probe approximation in the
literatures. The full description of the phase transition in such
systems needs to amount the affects of the backreacted fields on the
fix geometry. Also in this case and apart from the probe limit,
people find similar scalar condensation in the boundary
\cite{HartnollJHEP2008}-\cite{PanWang3}. All these calculations in
the probe limit or with back reactions have been done numerically,
based on the shooting method for solving the coupled of the
differential equations by appropriate boundary conditions. From the
analytical point of view of holographic superconductors, there are
three main methods in
literatures:\\
\emph{1-  The Sturm-Liouville  variational method} \cite{Siopsis}, which near the critical point, we replace
 the electromagnetic scalar gauge by an approximated
solution which satisfy the boundary conditions and then reduce the form of the scalar field equation to a usual
form of a self-adjoint equation in the functional theory of the real valued functions. By minimizing such
 functional by a suitable trial function one can find the best optimized critical temperature.
\emph{2-The
small parameter perturbation theory, in which we expand all the functions as a perturbative series with respect to the small parameter $\langle{\cal O}_{+}\rangle$ (it remains small just near the critical point $T=T_c$) and finds the corrections on the background or fields on it \cite{Herzog-2010,kanno}.}  \\
\emph{3-The Matching method. In this method, the asymptotic
solutions of fields near the AdS horizon to the horizon solutions at
an arbitrary (completely arbitrary) mid point are matched, and then
we find the expectation value of dual operator $\langle{\cal
O}_{+}\rangle$ and the
critical temperature $T_c$ analytically \cite{Gregory,Sebastiani}.}\\
Many models of the holographic superconductors have been
investigated  in four dimensions or higher-dimensions. However, it
is a possibility to consider a toy model for lower dimensional
holographic superconductors. Such lower dimensional models have been proposed for simplicity and also to find the nature of the phase transitions in the dilatonic black holes which have a very important role in the string theory. Another reason for consideration of a non realistic lower dimensional superconductors backs to the ability of the AdS/CFT dictionary in the reduction of the dimensions from $d>4$ to $2<d<4$.  Here you need to a lower dimension
AdS/CFT dictionary. The existence of such a dictionary depends upon
the string theory. But  in any case, we know that, for example such
theory works in  $AdS_{3}/CFT_{2}$ correspondence \cite{ren}.
Different aspects of such lower dimensional holographic superconductors have been discussed
in the probe limit effectively by the authors
\cite{ren},\cite{Bin1}. In addition, they showed that explicitly the
behaviors of the fields as the dynamical quantities near the phase
transition points. In the present work, we will use the variational
method to observe the analytical behavior of such a lower
dimensional holographic superconductors. Although people
investigated the problem before by applying the numerical
algorithms, however, here we will derive the
critical properties just be applying the analytical method. Firstly,
we review the main ideas of the holographic superconductors in such
lower dimensional systems. Generally, we know that now,
\cite{Gregory},\cite{Hart}, in the quantum theory on the boundary,
both dual operators correspond to the conformal dimension
$\Delta_{\pm}$, and explicitly they can be written as $m^2$, where
$m$ is mass of the scalar field. The work is planned as: in section
II, an $U(1)$ gauge field model along with the scalar field is
presented, within the background of a Banados-Teitelboim-Zanelli
(BTZ) like planar black hole; in section III, the phase transition
 is analytically investigated; in section
IV, we obtain the critical temperature $T_c$ versus the backreaction
up to the order $\kappa^2$. Finally we conclude in the last section.
\section{$1+1$ Holographic Superconductor}

The general lower dimensional, in fact $2+1$ gravitational bulk
action depicting a charged complex scalar field (we set Stuckelberg
field $\theta=0$, it means we take the scalar field read)
 with negative cosmological constant of Einstein-Maxwell action reads
\cite{ren}
\begin{eqnarray}\label{action}
S=\int d^3
x\sqrt{-g}\Big[\frac{1}{2\kappa^2}(R+\frac{2}{l^2})-\frac{1}{4}F^{ab}F_{ab}-|\nabla\zeta-i
e A\zeta |^2-m^2|\zeta|^2\Big].
\end{eqnarray}
Here, $\kappa$ is the usual three dimensional gravitational constant
$\kappa^2=8 \pi G_3$, $G_3$ is the Newton constant in three
dimensions and $g=|g_{\mu\nu}|$. We
need an AdS radius ~$l$  (non effective, because here we used just
the Einstein-Hilbert gravitational term), $e$  appears in the covariant derivative exactly
as a standard electric charge and $m=m_{\phi}$. We are interesting in
the effects of backreaction on the holographic superconductor. We
focus just on the s-wave cases. We must clarify the motivation of the s-wave approximation in holographic models of the type II high temperature superconductors. In the relativistic models of the gravity, we know that s-wave approximation is not a good approximation \cite{sergei1}.  s-wave refers to a scalar condensation not more. But you can have the Yang-Mills fields with $SU(2)$ symmetry which potentially they can generate another symmetry breaking of an axial vector type. The last case resembles the p-wave models.  So, just for more clarifications, we mention here that the two dimensional model of the superconductors which we proposed is a toy model and it will be more interesting that we can find a direct relation between this toy model and the results of a 4 dimensional model, by a principle like the detailed balance.\\
Another additional point is, here we study only the case of a single horizon and not multi horizon cases\cite{sergei2}. In the holographic set up for superconductors one must identify a temperature in his gravitational bulk model to the CFT temperature on the boundary because the partiotion function of bulk (stringy) equals to the partition function of the CFT. If the black hole has only one horizon, in this case, we can use the Hawking-Bekenstein (horizon) temperature as a reasonable candidate. But if our asymptotically AdS bulk has more than one horizon, for example, in the case of the charged BTZ like black holes, then we take the temperature of the real physical horizon (the temperature which is obtained by calculation the surface gravity of the biggest null hypersurface orthogonal surface) as the candidate for temperature of the CFT. In fact the effects of the quantum corrections and charged Maxwell field on the background of the bulk is very interesting problem and can be investigated in more details. Also, may be it become possible to relate the instability of such charged dilaton configurations in the AdS spacetime \cite{sergei3} to the symmetry breaking mechanism of the superconductors. The idea has motivation enough as a new work. Also, the effect of the charge in the dilatonic BTZ like black holes can produce the hyperscalling violations which it can modify the thermodynamics and the critical point of the second order phase transitions.\\
In this paper, by ignoring the quantum corrections or instabilities
for the gravity bulk sector, we use from an ansatz which is
described by the static spherically symmetric spacetime as
\begin{eqnarray}\label{ansatz}
ds^2=-f(\gamma)e^{-\beta(\gamma)}dt^2+\frac{d\gamma^2}{f(\gamma)}+\frac{\gamma^2}{l^2}dx^2~.
\end{eqnarray}
The $U(1)$ usual electromagnetic gauge field, $\gamma$ stands for
the radial coordinate, and the electromagnetic gauge is in the
following one form (in the language of differential geometry ) and
the scalar field
\begin{eqnarray}
A_t=a_0(\gamma)dt,\ \ \zeta\equiv\zeta(\gamma).
\end{eqnarray}
The scalar field phase is set to be zero, so it's natural that we
treat it as the real function. In the gravity sector of the bulk, we
must identify the temperature. The unique well behavior of this
static black hole is the Hawking-Unruh killing-Horizon temperature,
which it can be construed as the temperature of the dual CFT, This temperature's reading
may be is not a very trivial case, and can be calculated by
different methods, for example, by the method of the weak rotation
in the action to make it Euclidean or also by expansion of the
metric near the horizon and comparison of the resulting metric by
the Rindler geometry. By any method, the result for the horizon is
the same as
\begin{eqnarray}\label{temperature}
T=\left.\frac{f'(\gamma)e^{-\beta(\gamma)/2}}{4
\pi}\right|_{\gamma=\gamma_+}~.
\end{eqnarray}
To find the equations of motion, we remember that these equations
are the Maxwell equation for electromagnetic gauge and the
generalized Klein-Gordon equation, which we list them here
\begin{eqnarray}
\nabla_{\mu}F^{\mu\nu}=J^{\nu},\\
(D_\mu D^\mu-m^2)\zeta=0.
\end{eqnarray}
Here the current $J^{\nu}$ can be derived easily by the Noether theorem. The closed forms of the equations presented before as the following \cite{plb},
\begin{eqnarray}\label{eom}
\zeta ''(\gamma)+\zeta '(\gamma)
\left[\frac{1}{\gamma}+\frac{f'(\gamma)}{f(\gamma)}-\frac{\beta
'(\gamma)}{2}\right]+\zeta (\gamma) \left[\frac{e^2 \Phi (\gamma)^2
e^{\beta
(\gamma)}}{f(\gamma)^2}-\frac{m^2}{f(\gamma)}\right]&=&0~,\nonumber\\
a_0 ''(\gamma)+a_0 '(\gamma) \left[\frac{1}{\gamma}+\frac{\beta '(\gamma)}{2}\right]-\frac{2 e^2 a_0 (\gamma) \zeta (\gamma)^2}{f(\gamma)}&=&0~,\nonumber\\
f'(\gamma)+2 \kappa ^2 \gamma \left[\frac{e^2 a_0 (\gamma)^2 \zeta
(\gamma)^2 e^{\beta (\gamma)}}{f(\gamma)}+f(\gamma) \zeta
'(\gamma)^2+m^2 \zeta (\gamma)^2+\frac{1}{2}
 e^{\beta (\gamma)} a_0 '(\gamma)^2\right]-\frac{2 \gamma}{l^2}&=&0,\nonumber\\
\beta '(\gamma)+ 4 \kappa ^2 \gamma \left[\frac{q^2 a_0 (\gamma)^2
\zeta (\gamma)^2 e^{\beta (\gamma)}}{f(\gamma)^2}+\zeta
'(\gamma)^2\right]&=&0.
\end{eqnarray}
Here, $f'=\partial_{\gamma}f$. The
full numerical solutions to the given system is reported in
\cite{plb}.

Using the scaling symmetry, we can set the charge parameter equal to
$e=1$.
However, our approach is the analytical approach, and we will not
reproduce the wellknown results of the \cite{plb}. For this reason,
we will use the Sturm-Liouville (S-L) variational method. We have to think about two different
boundaries. When system stays in the normal phase, $\zeta(\gamma)=0$,
we find that the lapse function of the metric (redshift function)
$\beta$ is a constant and the analytic solutions to system
(\ref{eom}) lead to the well known exact black holes with the metric
coefficient and the potential function
\begin{eqnarray}
f(\gamma)=k+\frac{\gamma^2}{l^2}-\kappa^2\mu^2\log \gamma,\label{f0}\\
a_0(\gamma)=\rho+\mu\log \gamma.\label{Phi0}
\end{eqnarray}
Here $k=-\frac{\gamma_{+}^2}{l^2}+\kappa^2\mu^2\log \gamma_{+}$,
where, $\mu,\rho$ correspond to  the chemical potential charge density
in the dual field theory. Also about the (\ref{Phi0}), we mention here that for example, a Maxwell field in $AdS_3$ is also logarithmic, but is physical once appropriate counterterms
are added (and describes a vector operator that is not a conserved current). When $\kappa=0$, the metric coefficient
$f$ recovers the case of the Banados, Teitelboim, Zanelli(BTZ) black
hole. If we are interested to solve in superconducting stage, where
$\zeta\neq0$, we must locate the appropriate boundary conditions. We
must clarify these boundary conditions here. At the black hole
horizon $\gamma_+$, which is the positive real root of
$f(\gamma_+)=0$, all fields have regular
solutions\cite{plb}
\begin{eqnarray}\label{horizonboundry}
a_0(\gamma_+)=0,\ \
\zeta'(\gamma_+)=\frac{m^2}{f'(\gamma_+)}\zeta(\gamma_+),
\end{eqnarray}
and the metric ansatz satisfy \cite{plb}
\begin{eqnarray}\label{horizonmetric}
f'(\gamma_+)&=&\frac{2 \gamma_+}{l^2}-2 \kappa ^2 \gamma_+ \left[m^2
\zeta (\gamma_+)^2+\frac{1}{2}
e^{\beta (\gamma_+)} a_0 '(\gamma_+)^2\right],\nonumber\\
\beta '(\gamma_+)&=&-4 \kappa ^2 \gamma_+ \left[\frac{e^2 a_0'
(\gamma_+)^2 \zeta (\gamma_+)^2 e^{\beta
(\gamma_+)}}{f'(\gamma_+)^2}+\zeta '(\gamma_+)^2\right].
\end{eqnarray}
Far from the horizon boundary, at the spatial infinity which it
coincides exactly on the AdS boundary, the asymptotic performance of
the solutions is
\begin{eqnarray}
&&\beta\rightarrow 0,\ \ f(\gamma)\sim \frac{\gamma^2}{l^2}\nonumber\\
&&a_0(\gamma)\sim \mu \log \gamma,\ \ \zeta(\gamma)\sim
\frac{\zeta_-}{\gamma^{\Delta-}}+\frac{\zeta_+}{\gamma^{\Delta+}},
\end{eqnarray}
where as the usual exponent $\Delta_\pm$ denote the conformal dimension. For $-1\leq m^2<0$, both the fields are
normalizable, so, the boundary conditions are applied in such a way
that one just becomes zero. After applying these appropriate
boundary conditions that either $\zeta_-$ or $\zeta_+$ becomes
extinct, a one parameter family of solutions is obtained using the
shooting method algorithms \cite{plb}.
\section{Analytical investigation of the holographic superconductors}

The S-L method \cite{Siopsis} is used here for the analytical
investigation of the properties of  holographic
superconductor  with backreactions. Further, the
relation between the critical temperature $T_c$ and charge density
$\rho$ will be derived near the phase transition point and resolve
the backreaction effects. As the usual, it's adequate to introduce a
new variable $z = \frac{\gamma_+}{\gamma}$, so, the Einstein,
Maxwell and the scalar equations can be written as

\begin{eqnarray}
\zeta ''(z)+\frac{\zeta '(z)}{z}
\left[1+\frac{zf'(z)}{f(z)}-\frac{z\beta
'(z)}{2}\right]+\frac{\gamma_{+}^2\zeta (z)}{z^4} \left[\frac{e^2
a_0 (z)^2 e^{\beta
(z)}}{f(z)^2}-\frac{m^2}{f(z)}\right]&=&0~,\label{Psiz}\\
a_0 ''(z)+\frac{a_0 '(z)}{z} \left[1-\frac{z\beta '(z)}{2}\right]-\frac{2 e^2\gamma_{+}^2 a_0 (z) \zeta (z)^2}{z^4f(z)}&=&0~,\label{phiz}\\
f'(z)-2 \kappa ^2 \frac{\gamma_{+}^2}{z^3} \left[\frac{e^2 a_0
(z)^2 \zeta (z)^2 e^{\beta (z)}}{f(z)}+\frac{f(z) \zeta
'(z)^2z^4}{\gamma_{+}^2}+m^2 \zeta (\gamma)^2+\frac{1}{2}
 \frac{e^{\beta (z)} a_0 '(z)^2z^4}{\gamma_{+}^2}\right]-\frac{2 \gamma_{+}^2}{l^2z^3}&=&0,\label{fz}\\
\beta '(z)- 4 \kappa ^2 \frac{\gamma_{+}^2}{z^3} \left[\frac{e^2 a_0
(z)^2 \zeta (z)^2 e^{\beta (z)}}{f(z)^2}-\frac{z^4\zeta
'(z)^2}{\gamma_{+}^2}\right]&=&0\label{chiz}.
\end{eqnarray}
where, in this new system, $f'=\frac{df}{dz}$.\\
Following the perturbation scheme in the \cite{kanno}, since the
value of the scalar operator $< O_{+} >$ (or $< O_{-} >$) is small
close to the critical point, it can be introduced as an expansion
parameter
\begin{eqnarray}
\epsilon\equiv< O_{i} >,\ \  i= +,-.
\end{eqnarray}

Note that, in the perturbation method and close to the critical
point, our interest is in the solutions for which the scalar field
$\zeta$ is small, therefore from equations (\ref{Psiz}-\ref{chiz}) we
must extend the gauge field $a_0$, the scalar field $\zeta$,  and
the metric functions $f(z),\beta(z)$ as
\begin{eqnarray}
\zeta=\Sigma_{k=1}^{\infty}\epsilon^k\zeta_k,\ \ a_0=\Sigma_{k=0}^{\infty}\epsilon^{2k}a_{0(2k)},\\
f=\Sigma_{k=0}^{\infty}\epsilon^{2k}f_{2k},\ \ \beta=\Sigma_{k=1}^{\infty}\epsilon^{2k}\beta_{2k}.
\end{eqnarray}
where the metric function $f(z)$ and $\beta(z)$  can be expanded around the BTZ spacetime which are the exact solutions in the probe limits. Also for the chemical potential $\mu$, we will allow it to be expanded as the following series form
\begin{eqnarray}
\mu=\Sigma_{k=0}^{\infty}\epsilon^{2k}\delta\mu_{2k}.
\end{eqnarray}
where at least  $\delta\mu_{2}> 0$. Thus, close to the phase
transition,
\begin{eqnarray}
\epsilon\approx\Big(\frac{\mu-\mu_0}{\delta\mu_{2}}\Big)^{\frac{1}{2}}\label{exponent},
\end{eqnarray}
whose critical exponent from the quantum holographic picture
$\beta=\frac{1}{2}$ as a result of mean field
theory . Obviously, as observed, the order
parameter vanishes and phase
transition can happen if $\mu\rightarrow\mu_0$, which shows $\mu$ is $\mu_c=\mu_0$.\\
At the zeroth order, we can get the solution $a_0$ from
(\ref{phiz}), i.e., the electromagnetic field behaves like
$a_0(z)=\rho+\mu\log\frac{\gamma_{+}}{z}$, which by applying the
boundary condition $a(\gamma_{+})=0$ gives a relation
$\mu_0=-\frac{\rho}{\log \gamma_{+}}$.
 At the critical
point $\mu_c$, we can find $\mu_0=\mu_c=-\frac{\rho}{\log
\gamma_{+c}}$, where $\gamma_{+c}$ is the radius of the horizon at
the critical point. For employing the analytical S-L method, we will
set
\begin{eqnarray}
a_0=-\lambda \gamma_{+c}\log z,\ \
\lambda=-\frac{\rho}{\gamma_{+c}\log \gamma_{+c}}.
\end{eqnarray}
Since $\zeta=0$, by inserting this solution into  (\ref{fz}), we
obtain the metric function in the probe limit
\begin{eqnarray}
f_0(z)=\gamma_{+}^2g(z)=\gamma_{+}^2\Big(\kappa^2\lambda^2\log
z-\frac{1}{l^2z^2}+\frac{1}{l^2}\Big).
\end{eqnarray}
Here, we define a new function $g(z)$ for simplicity in the
following calculation, using the boundary conditions on the horizon
$f_0(z=1)=0$, the constant term $\frac{1}{l^2}$ is obtained. Now in
the first order approximation, the asymptotic AdS solution for
$\zeta_1$ can be expressed as
\begin{eqnarray}
\zeta_1(z)\sim \frac{\zeta_-
}{\gamma_{+}^{\Delta-}}z^{\Delta-}+\frac{\zeta_+
}{\gamma_{+}^{\Delta+}}z^{\Delta+}, \ \
\zeta_{\pm}=\langle{O_{\pm}}\rangle.
\end{eqnarray}

So, introducing a variational completely trial function $F(z)$ close
to $z = 0$

\begin{eqnarray}
\zeta_1(z)\sim\frac{<O_i>}{\gamma_{+}^{\Delta_{i}}}z^{\Delta_{i}}F(z)\label{Psi1}.
\end{eqnarray}
where, $F(0) = 1$ and $F'(0) = 0$. Substituting equation
(\ref{Psi1}) into equation (\ref{Psiz}), we get,

\begin{eqnarray}
F''+\Big[\frac{2\Delta_{i}+1}{z}+\frac{g'}{g}\Big]F'+\Big[\frac{\lambda^2(\log z)^2}{z^4g^2}-\frac{m^2}{z^4g}+\frac{\Delta_{i}}{z^2}(\Delta_{i}+\frac{zg'}{g})\Big]F=0\label{F}.
\end{eqnarray}

We can convert  (\ref{F}) to be
\begin{eqnarray}
(TF')'+T\Big[U+\lambda^2 V\Big]F=0\label{FSL},
\end{eqnarray}
with
\begin{eqnarray}
T=gz^{2\Delta_{i}+1},\ \
U=-\frac{m^2}{z^4g}+\frac{\Delta_{i}}{z^2}(\Delta_{i}+\frac{zg'}{g}),\
\ V=\frac{(\log z)^2}{z^4g^2}.
\end{eqnarray}
From the SL eigenvalue problem in the real valued functional theory,
the expression to minimize eigenvalue of $\lambda^2$ is
\begin{eqnarray}
\lambda^2=\frac{\int_{0}^{1}T(F'^2-UF^2)dz}{\int_{0}^{1}TVF^2dz}\label{lambda1}.
\end{eqnarray}
Note that, here we are interesting just to the corrections of the backreaction term, i.e. we want to obtain the $\lambda^2$ up to the order of $\kappa^2$. So it is useful to write the functions
\begin{eqnarray}
T=T_0(z)+\kappa^2\lambda^2T_1(z),\ \ T_1(z)=z^{2\Delta_{i}+1}\log z ,\  \ T_0(z)=\frac{z^{2\Delta_{i}+1}}{l^2}(1-z^{-2}),\\
g=g_0(z)+\kappa^2\lambda^2g_1(z),\ \ g_1(z)=\log z,\ \ g_0(z)=\frac{1}{l^2}(1-z^{-2}),\\
U=U_0(z)+\kappa^2\lambda^2 U_1(z),\ \ U_0(z)=-\frac{m^2}{z^4g_0(z)}+\frac{\Delta_{i}}{z^2}(\Delta_{i}+\frac{zg'_0(z)}{g_0(z)}),    \ \ U_1(z) = \frac{m^2g_1(z)}{z^4g_0(z)^2} +\frac{\Delta_{i}}{z^3} \Big(\frac{g_1(z)}{g_0(z)}\Big)',\\
V=V_0(z)+\kappa^2\lambda^2 V_1(z),\ \ V_0(z)=\frac{(\log z)^2}{z^4g_0(z)^2}    ,\ \ V_1(z) =\frac{(\log z)^2g_1(z)}{z^4g_0(z)^3}.
\end{eqnarray}
So, the (\ref{lambda1}) revised as
\begin{eqnarray}
\lambda^2\approx\lambda^2+\kappa^2\lambda_1^2+O(\kappa^4),\\
\lambda^2=\frac{\alpha_1}{\alpha_2},\ \
\lambda_1^2=\frac{\alpha_1}{\alpha_2^3}(\alpha_2\beta_1-\alpha_1\beta_2)
\label{lambda2},
\end{eqnarray}
where
\begin{eqnarray}
\alpha_1=\int_{0}^{1}T_0(z)(F'^2-U_0(z)F^2)dz,\label{a}\\
\beta_1=\int_{0}^{1}\Big[T_1(z)(F'^2-U_0(z)F^2)-F^2T_0(z)U_1(z)\Big]dz,\label{b}\\
\alpha_2=\int_{0}^{1}T_0(z)V_0(z)F^2dz,\label{c}\\
\beta_2=\int_{0}^{1} F^2(T_0(z)V_1(z)+T_1(z)V_0(z))dz\label{d}.
\end{eqnarray}
The first term in (\ref{lambda2}) is the probe limit ($\kappa^2=0$) value for the
eigenvalue while the first order correction of the back reaction $\kappa^2$, is shown in second term.\\
In the following section, the analytical results are presented in
solving (\ref{lambda2}) for several values of the backreaction
$\kappa^2$ with $m^2 = 0 $ to compare with the numerical results in
\cite{plb}. We will use  $F(z)=1-az^2$.

\section{The critical temperature $T_c$}

When we are choosing the massless scalar field, the conformal
dimension $\Delta_{i}$ takes the form
\begin{eqnarray}
\Delta_{i}=\Delta_{+}=2.
\end{eqnarray}
For describing the condensate, we choose only $\zeta_{+}$ dual to the
scalar operator in the boundary field theory. So in this case the
(\ref{a}-\ref{d}) read as
\begin{eqnarray}
\alpha_1&=&-1 + 1.33 a - 0.67 a^2,\label{aa}\\
\beta_1&=&0.25 + 0.28a - 0.22 a^2 ,\label{bb}\\
\alpha_2&=&-0.0505142 + 0.0385285a - 0.010005 a^2 ,\label{cc}\\
\beta_2&=&-0.0280578 + 0.0263808a - 0.00758224 a^2 \label{dd}.
\end{eqnarray}So,  for (\ref{lambda2}) we have
\begin{eqnarray}
\lambda^2&=&\frac{\Sigma_{n=0}^7 c_n a^n}{(43.6443 - 33.2886 a + 8.64429 a^2)^2},\\
c_0&=&746496- 77.5004 \kappa^2,\\
 c_1&=&-1.99066\times10^6 + 231.354 \kappa^2,\\
 c_2&=&2.32243\times10^6 - 328.431 \kappa^2,\\
 c_3&=&-1.3271\times10^6 + 287.341 \kappa^2,\\
 c_4&=&331776. - 168.438 \kappa^2,\\
 c_5&=&67.323 \kappa^2,\\
 c_6&=&-17.8282 \kappa^2,\\
 c_7&=&2.85173 \kappa^2.
\end{eqnarray}

We list the $\lambda^2_{Min}$ with the chosen strength of the
backreaction $\kappa$ for the condensates of the scalar operator
$<O_{+}>$ 3-dimensional case of hairy black hole
background\footnote{ The case with $m^2=-1$ does  not have
convergence.}.

\begin{center}
\begin{table}[ht]
\caption{\label{Table.IV} The dependence of the eigenvalue
$\lambda^2_{Min}$ and the optimal value of the parameter $a$  on the
backreaction $\kappa^2$ with $m^2=0$. }
\begin{tabular}{|c|c|c|c|c|c|}
\hline $\kappa^{2}$ & 0  &  0.05  &  0.1  &  0.15  &  0.2 \\ [0.5ex]
\hline
 $a$ &~0.759109~&~0.759108~&~0.759107~&~0.759106~&~0.759105~\\ [0.5ex]
\hline
 $\lambda^2_{Min}$ &~189.394~&~187.312~&~179.392~&~169.216~&~158.125~\\ \hline
\end{tabular}
\end{table}
\end{center}

Now, critical temperature $T_c$ can be taken for different values of
the backreaction $\kappa$ and from the following relation
\begin{eqnarray}
T_c=\frac{\gamma_{+c}}{4\pi}-\frac{\kappa^2\mu^2}{4\pi
\gamma_{+c}},\ \ \mu=\mu_c=-\frac{\rho}{\log \gamma_{+c}}.
\end{eqnarray}
Finally we have,
\begin{eqnarray}
\frac{T_c}{\mu_c}=\frac{1}{4\pi \lambda_{Min}}(1-\kappa^2\lambda_{Min}^2).
\end{eqnarray}
So we can put the data in the following table.  We observe that the
analytic results for the critical temperature are consistent with
the numerical result \cite{plb}.

\begin{center}
\begin{table}[ht]
\caption{\label{Table.IV} The dependence of the critical temperature
$\frac{T_c}{\mu_c}$  on the backreaction $\kappa^2$. Obviously,
growth of back reaction decrease the critical temperature, with
$m^2=0$.}
\begin{tabular}{|c|c|c|c|c|c|}
\hline $\kappa^{2}$ & 0  &  0.05  &  0.1  &  0.15  &  0.2 \\ [0.5ex]
\hline
 $\frac{T_c}{\mu_c}$ &~0.0578239~&~0.0489749~&~0.0193732~&~0.0158489~&~0.01313246~\\ [0.5ex]
\hline
\end{tabular}
\end{table}
\end{center}

\begin{figure*}[thbp]
\begin{tabular}{rl}
\includegraphics[width=7.5cm]{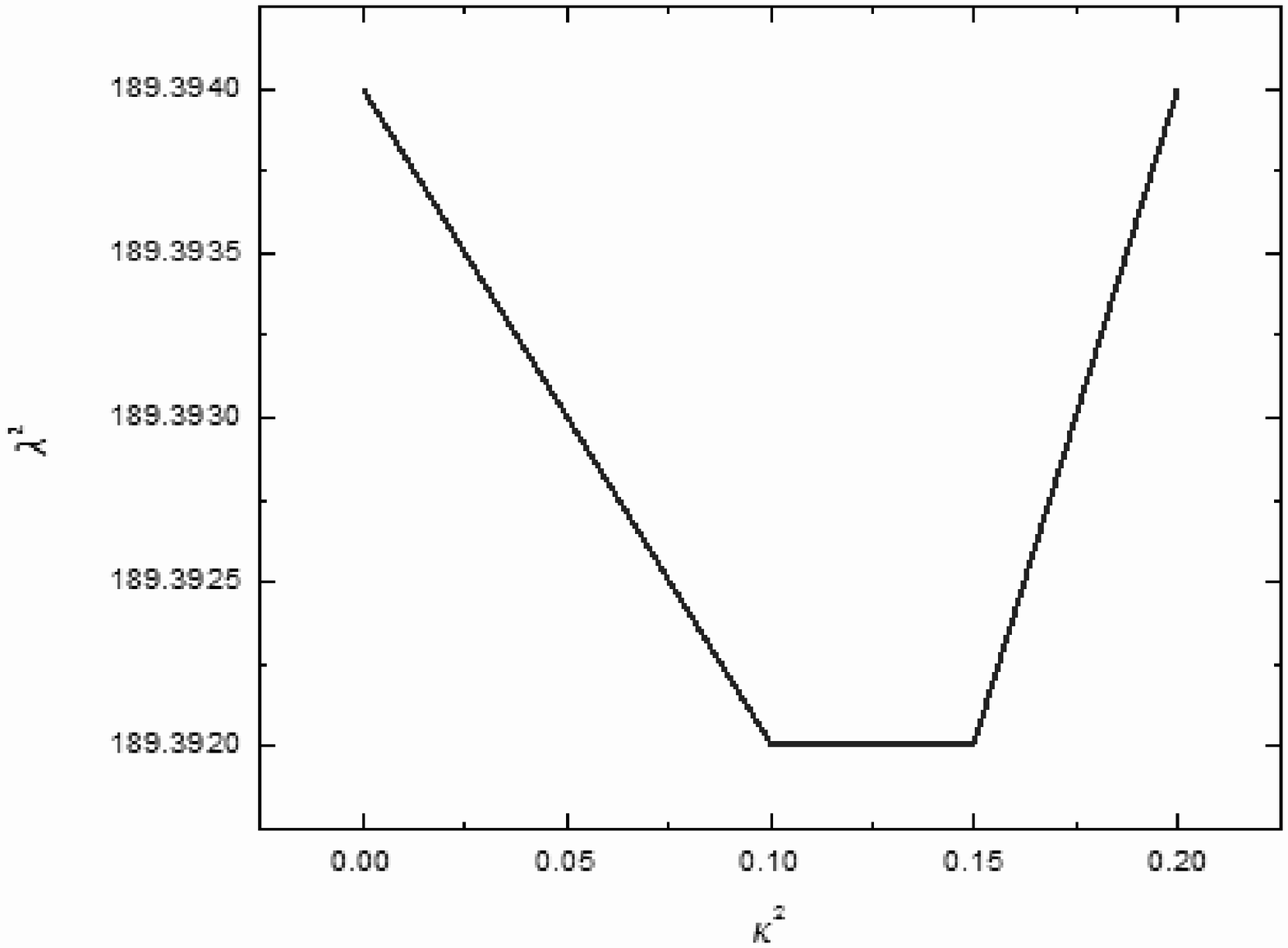}&
\includegraphics[width=7.5cm]{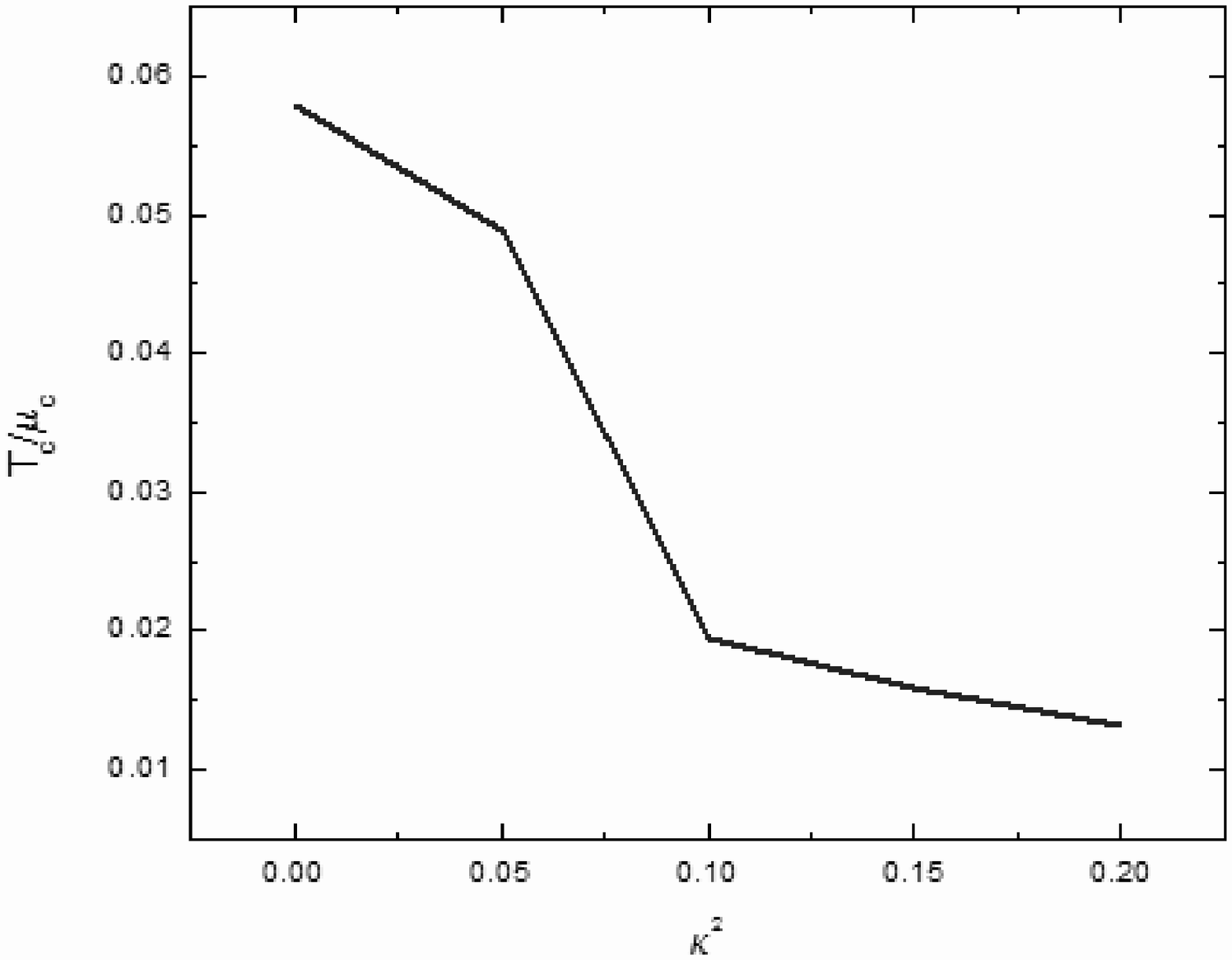} \\
\end{tabular}
\caption{ (\textit{Left}) The curve depicts the eigen value
$\lambda^2$ as a function of the $\kappa^2$.
   (\textit{Right}) Variation of the $\frac{T_c}{\mu_c}$ versus $\kappa^2$. It shows that the backreaction decreases  $T_c$.  }
\end{figure*}

\section{Conclusions and discussions}

We investigated the holographic superconductors in the three dimensional
gravitational bulk background with backreactions.We have used the Sturm–Liouville
eigenvalue variational method for the analytical investigation of the holographic super conductor’s
properties with backreactions. We found that in the fully backreacted BTZ spacetime,
our analytical results shows a very good coincidence with numerically computed results.
According to our analytical results, the backreaction decrease the critical temperature
of the superconductor, which can be used to support the numerical results that the condensation
can be hindered by the backreactions.




\end{document}